\documentclass[aps,prl,twocolumn,groupedaddress]{revtex4}

\usepackage{graphicx}
\usepackage{dcolumn}
\usepackage{bm}

\begin{document}


\title{Comment on ``Oxygen as a Site Specific Probe of the Structure of Water and Oxide Materials'', PRL {\textbf 107}, 144501 (2011)}


\author{A.K.Soper}
\affiliation{STFC Rutherford Appleton Laboratory, ISIS Facility, Harwell Oxford, Didcot, Oxfordshire, OX11 0QX, UK}
\email[]{alan.soper@stfc.ac.uk}

\author{C.J.Benmore}
\affiliation{Argonne National Laboratory, 9700 S. Cass Ave., Argonne, Il 60439}


\date{\today}



\maketitle


A recent paper by Zeidler \textit{et al.} \cite{zeidlerwater} describes a neutron scattering experiment on water in which oxygen isotope substitution is successfully achieved for the first time. Differences between scattering patterns with different oxygen isotopes give a combination of the O-O and O-H (or O-D) structure factors, and the method elegantly minimises some of the problematic inelasticity effects associated with neutron scattering from hydrogen. Particular conclusions of the new work are that the OH bond length in the H$_2$O molecule is about 0.005\AA\ longer than the same bond in D$_2$O, and that the hydrogen bond peaks in both liquids are at about the same position. 

Notwithstanding the substantial progress demonstrated by the new work, the comparison with our own results \cite{soper2008} by Zeidler \textit{et al.} is in our opinion misleading. The last paragraph of \cite{zeidlerwater} states that the new experimental results ``...\textit{originate directly from measured data sets and not from a refinement of models using the diffraction data, wherein different starting points can lead to quite different conclusions [36,41].}'' Firstly the results reported in references 36 \cite{soper2008} and 41 \cite{soper2000} do indeed originate directly from measured datasets, as is abundantly clear from the papers themselves. Equally the reported OH bond lengths in \cite{soper2008} are a direct consequence of the particular diffraction data presented, as highlighted in Fig. 2 of that paper, and do not depend on the computer simulation method used. Both papers use different empirical potential structure refinement (EPSR) computer simulations to fit radiation scattering data and derive OO, OH and HH radial distribution functions compatible with those data. Comparing Fig 3. of \cite{soper2008} with Fig. 6 of \cite{soper2000} and Fig. 6 of \cite{soper2007}, which include independent datasets from both x-ray and neutron (reactor and pulsed) sources, and from ``different starting points'', the radial distribution functions for water extracted by this method from all these different datasets are in fact very similar. But they are of course not identical and this is known to be due to systematic uncertainties which are difficult to quantify, particularly where scattering from hydrogen or small isotope differences is involved \cite{fischer2006}. Zeidler \textit{et al.} distance themselves from the EPSR approach, yet by choosing one potential, TTM3-F, out of several possibilities they have in fact performed what EPSR does automatically, namely selected a potential that reproduces their data. This \textit{simulation} is then used to claim that the hydrogen bond length in H$_2$O is about the same length as that in D$_2$O, just as in \cite{soper2008} the EPSR simulation is used to claim these bonds are different. Note however that the results \cite{soper2008} reproduce the change in x-ray scattering pattern between heavy and light water \cite{hart2005}, something that TTM potentials tend to underestimate \cite{paesani2009}.  

\begin{figure}
\includegraphics[width=0.48\textwidth]{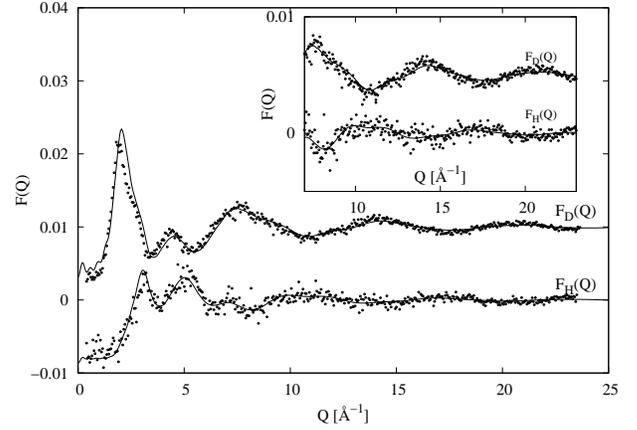}
\caption{\label{fig1} Simulated first order difference functions (solid lines), $\Delta F_{D}(Q)= 0.0059[S_{OD}(Q)-1]+0.00262[S_{OO}(Q)-1]$(top), $\Delta F_{H}(Q)= -0.0033[S_{OH}(Q)-1]+0.00263[S_{OO}(Q)-1]$(bottom), based on the EPSR simulations reported in \cite{soper2008}. The oxygen isotope substitution data from \cite{zeidlerwater} are shown as open circles. The inset shows the high $Q$ region in more detail. Curves are shifted vertically for clarity.} 
\end{figure}

To illustrate the ambiguities associated with using any particular dataset to characterise the structure of water we have constructed the oxygen first order difference functions, $\Delta F_{H}(Q)$ and $\Delta F_{D}(Q)$ (Figure 1(a) of \cite{zeidlerwater}) using the simulated OO and OH partial structure factors that were extracted in the previous analysis \cite{soper2008}, weighted according to the new values of the oxygen isotope scattering lengths, and without refinement against the new data (Figure \ref{fig1}). Although there is some disagreement in the region of the main peak near $Q=2$\AA$^{-1}$ in the $\Delta F_{D}(Q)$ function, it is striking  how close the earlier EPSR simulations approach the new data. A similar discrepancy in $\Delta F_{D}(Q)$ between simulation and experiment near $Q=2$\AA$^{-1}$ is observed in Figure 1(a) of \cite{zeidlerwater}. 
  
\begin{thebibliography}{7}
\expandafter\ifx\csname natexlab\endcsname\relax\def\natexlab#1{#1}\fi
\expandafter\ifx\csname bibnamefont\endcsname\relax
  \def\bibnamefont#1{#1}\fi
\expandafter\ifx\csname bibfnamefont\endcsname\relax
  \def\bibfnamefont#1{#1}\fi
\expandafter\ifx\csname citenamefont\endcsname\relax
  \def\citenamefont#1{#1}\fi
\expandafter\ifx\csname url\endcsname\relax
  \def\url#1{\texttt{#1}}\fi
\expandafter\ifx\csname urlprefix\endcsname\relax\def\urlprefix{URL }\fi
\providecommand{\bibinfo}[2]{#2}
\providecommand{\eprint}[2][]{\url{#2}}

\bibitem[{\citenamefont{Zeidler et~al.}(2011)\citenamefont{Zeidler, Salmon,
  Fischer, Neuefeind, Simonson, Lemmel, Rauch, and Markland}}]{zeidlerwater}
\bibinfo{author}{\bibfnamefont{A.}~\bibnamefont{Zeidler}},
  \bibinfo{author}{\bibfnamefont{P.~S.} \bibnamefont{Salmon}},
  \bibinfo{author}{\bibfnamefont{H.~E.} \bibnamefont{Fischer}},
  \bibinfo{author}{\bibfnamefont{J.~C.} \bibnamefont{Neuefeind}},
  \bibinfo{author}{\bibfnamefont{J.~M.} \bibnamefont{Simonson}},
  \bibinfo{author}{\bibfnamefont{H.}~\bibnamefont{Lemmel}},
  \bibinfo{author}{\bibfnamefont{H.}~\bibnamefont{Rauch}}, \bibnamefont{and}
  \bibinfo{author}{\bibfnamefont{T.~E.} \bibnamefont{Markland}},
  \bibinfo{journal}{Phys. Rev. Lett.} \textbf{\bibinfo{volume}{107}},
  \bibinfo{pages}{145501} (\bibinfo{year}{2011}).

\bibitem[{\citenamefont{Soper and Benmore}(2008)}]{soper2008}
\bibinfo{author}{\bibfnamefont{A.~K.} \bibnamefont{Soper}} \bibnamefont{and}
  \bibinfo{author}{\bibfnamefont{C.~J.} \bibnamefont{Benmore}},
  \bibinfo{journal}{Phys. Rev. Lett.} \textbf{\bibinfo{volume}{101}},
  \bibinfo{pages}{065502} (\bibinfo{year}{2008}).

\bibitem[{\citenamefont{Soper}(2000)}]{soper2000}
\bibinfo{author}{\bibfnamefont{A.~K.} \bibnamefont{Soper}},
  \bibinfo{journal}{Chem. Phys.} \textbf{\bibinfo{volume}{258}},
  \bibinfo{pages}{121 } (\bibinfo{year}{2000}).

\bibitem[{\citenamefont{Soper}(2007)}]{soper2007}
\bibinfo{author}{\bibfnamefont{A.~K.} \bibnamefont{Soper}},
  \bibinfo{journal}{J. Phys. Condens. Matter} \textbf{\bibinfo{volume}{19}},
  \bibinfo{pages}{335206} (\bibinfo{year}{2007}).

\bibitem[{\citenamefont{Fischer et~al.}(2006)\citenamefont{Fischer, Barnes, and
  Salmon}}]{fischer2006}
\bibinfo{author}{\bibfnamefont{H.~E.} \bibnamefont{Fischer}},
  \bibinfo{author}{\bibfnamefont{A.~C.} \bibnamefont{Barnes}},
  \bibnamefont{and} \bibinfo{author}{\bibfnamefont{P.~S.}
  \bibnamefont{Salmon}}, \bibinfo{journal}{Rep. Progr. Phys.}
  \textbf{\bibinfo{volume}{69}}, \bibinfo{pages}{233 } (\bibinfo{year}{2006}).

\bibitem[{\citenamefont{Hart et~al.}(2005)\citenamefont{Hart, Benmore,
  Neuefeind, Kohara, Tomberli, and Egelstaff}}]{hart2005}
\bibinfo{author}{\bibfnamefont{R.~T.} \bibnamefont{Hart}},
  \bibinfo{author}{\bibfnamefont{C.~J.} \bibnamefont{Benmore}},
  \bibinfo{author}{\bibfnamefont{J.}~\bibnamefont{Neuefeind}},
  \bibinfo{author}{\bibfnamefont{S.}~\bibnamefont{Kohara}},
  \bibinfo{author}{\bibfnamefont{B.}~\bibnamefont{Tomberli}}, \bibnamefont{and}
  \bibinfo{author}{\bibfnamefont{P.~A.} \bibnamefont{Egelstaff}},
  \bibinfo{journal}{Phys. Rev. Lett.} \textbf{\bibinfo{volume}{94}},
  \bibinfo{pages}{047801} (\bibinfo{year}{2005}).

\bibitem[{\citenamefont{Paesani and Voth}(2009)}]{paesani2009}
\bibinfo{author}{\bibfnamefont{F.}~\bibnamefont{Paesani}} \bibnamefont{and}
  \bibinfo{author}{\bibfnamefont{G.~A.} \bibnamefont{Voth}},
  \bibinfo{journal}{J. Phys. Chem. B} \textbf{\bibinfo{volume}{113}},
  \bibinfo{pages}{5702 } (\bibinfo{year}{2009}).

\end{thebibliography}
\end{document}